\documentclass[superscriptaddress,secnumarabic,
amssymb,amsmath,nobibnotes,aps,prd,nofootinbib]{revtex4}
\usepackage{graphicx}
\usepackage{epsf}
\usepackage{bm,color}
\usepackage{float}
\usepackage{amsmath}
\usepackage{amsfonts}
\usepackage{amssymb}
\usepackage{epstopdf}
\usepackage{natbib}%
\usepackage{subcaption}
\usepackage{hyperref}
\hypersetup{
	bookmarks=true,         
	unicode=false,          
	pdftoolbar=true,        
	pdfmenubar=true,        
	pdffitwindow=false,     
	pdfstartview={FitH},    
	pdftitle={MHI},    
	pdfauthor={Author},     
	pdfsubject={Subject},   
	pdfcreator={Creator},   
	pdfproducer={Producer}, 
	pdfkeywords={keyword1} {key2} {key3}, 
	pdfnewwindow=true,      
	colorlinks=true,       
	linkcolor=red,          
	citecolor=cyan,        
	filecolor=magenta,      
	urlcolor=blue,           
	linktocpage=true
}
\providecommand{\U}[1]{\protect\rule{.1in}{.1in}}
\newcommand{\be}{\begin{equation}}
\newcommand{\ee}{\end{equation}}

\newcommand{\mincir}{\raise
	-3.truept\hbox{\rlap{\hbox{$\sim$}}\raise4.truept\hbox{$<$}\ }}
\newcommand{\magcir}{\raise
	-3.truept\hbox{\rlap{\hbox{$\sim$}}\raise4.truept\hbox{$>$}\ }}

\def\bea{\begin{eqnarray}}
\def\eea{\end{eqnarray}}
\def\ba{\begin{array}}
	\def\ea{\end{array}}

\def\beq{\begin{equation}}
\def\eeq{\end{equation}}

\newcommand{\eq}{Eq.\eqref}
\newcommand{\fig}{Fig.\ref}
\begin{document}
	\title{Returning Back to Mukhanov Parametrization of Inflationary Equation of State}
	\author{Barun Kumar Pal}
	\email{terminatorbarun@gmail.com}
	\affiliation{Netaji Nagar College For Women, Kolkata-700092, West Bengal, India}
	\begin{abstract}
		We have re-examined Mukhanov parametrization for inflationary equation of state,  $1+\omega=\frac{\beta}{({N}+1)^\alpha}$,  in the light of Planck 2018 results and latest bound of tensor-to-scalar ratio employing Hamilton-Jacobi formalism. 	
		We have found that the current observational values of scalar spectral index and tensor-to-scalar ratio can be used efficiently to constrain the model parameters.  The recent bound of $r<0.032$ has been used to put an upper bound on one of the model parameter. Whereas   1-$\sigma$ bound of the scalar spectral index $0.9607\leq n_{_S}\leq 0.9691$ along with the upper bound of tensor-to-scalar ratio provided  restriction on the other model parameter $1.50<\alpha\leq2.20$. These bounds however depend on the number of e-foldings still left before the end of inflation and whenever $1.50<\alpha\leq2.20$  we can find appropriate values of the other model parameter $\beta$ so that the observational predictions are in tune with the latest available inflationary observables.   
		We have further utilized the predictions from forthcoming CMB missions in the likes of CMB-S4 and LiteBIRD in order to obtain bounds on the model parameters. We find that detection of gravity waves would help  us  constrain the model parameters further.  But in the absence of detection of primordial gravity wave signal  by these CMB missions may rule out Mukhanov parametrization. 
		
	\end{abstract}

	\maketitle

\section{Introduction}\label{sec1}
Cosmic inflation, by far the best bet for early universe scenario, has started its journey more than four decades ago in order to figure out the Big Bang puzzles. It has resolved the problems of Big Bang theory in a very elegant way. Not only that, inflation has surprised us by supplying  quantum seeds for cosmological fluctuations observed in the large scale structure of matter and in cosmic microwave background radiation. Unfortunately due to the lack of specific compelling model of inflation, it is still a paradigm. During its  long existence plenty of inflationary models have been proposed. But  the situation has changed drastically since the availability of precise observational data.  The finespun data of late \cite{akrami2020planck, aghanim2020planck, ade2014planck, spergel2007}  has  ruled out many inflationary models already. The future CMB missions in the likes of BICEP2/Keck \cite{ade2018constraints}, CMB-S4 \cite{abazajian2016cmb}, LiteBird \cite{matsumura2014mission} will further reduce the number of viable  inflationary models.  

The inflationary paradigm is  explored mostly through the avenue of defining a particular potential for the inflaton, which may be of field theoretic origin or may be purely phenomenological. The advantage of this approach is that we can directly establish link with the underlying physics behind cosmic inflation. There are other alternative approaches which may not have direct field theoretic motivation but may have viable observational predictions \cite{vagnozzi2022challenge}. The equation-of-state formalism is one such alternative approach, first introduced in Ref.\cite{mukhanov2013quantum} and later discussed in  several works \cite{martin2014encyclopaedia, roest2014universality,garcia2014large,barranco2014model,bamba2014inflationary, bamba2014reconstruction, boubekeur2015phenomenological,gariazzo2017primordial, baruneos2023}. With two free parameters equation-of-state formalism is able to mimic recent observations.  Though originally equation of state formalism  has been advertised as the model independent  way to investigate inflationary paradigm. But in reality we shall see that specifying an equation of state  boils down to choosing a particular potential for inflation \cite{martin2016observational}. Nonetheless, since a specific model of inflation is yet to be separated out from the spectrum of possible models and alternatives, equation-of-state formalism renders an elegant way to study inflationary paradigm  as, in principle, it  can encompass many models of inflation. Also in this method  the inflationary predictions can be easily confronted with observational data without going into the details of early universe physics.

The dynamical equations of motion for inflation do not, in general, possess analytical solutions. We need to employ  approximation technique to analyze   inflationary dynamics and the most widely used technique is the slow-roll approximation \cite{linde1982, albrecht1982, liddle1994}. But slow-roll is not the only method for successful implementation of inflationary models and solutions beyond slow-roll approximations has also been found \cite{wands1996}. In order to incorporate different types of inflationary models irrespective of slow-roll approximations, Hamilton-Jacobi formalism \cite{muslimov1990, salopek1990} has turned out to be an excellent tool. In this formulation the Hubble parameter, $H$, is treated as the fundamental quantity in contrast to the traditional approach of inflationary cosmology where we need  to specify a particular form of the scalar filed potential, $V$. Hamilton-Jacobi formulation allows us to tract the scalar field evolution in an exact way. The main advantage of Hamilton-Jacobi formalism  is that it is more accurate than the usual slow-roll method as it also takes into account the effect from  the kinetic term present in inflationary dynamics. One of the main drawback of this formulation is that it breaks down if the field undergoes oscillations \cite{lidsey1997}.

Nowadays scalar spectral index has come out as the main eliminator between different inflationary models along with the tensor-to-scalar ratio. With the increase in precision level of present day detectors many inflationary models are expected to be wiped out in near future.  In this article we have re-investigated Mukhanov parametrization employing Hamilton-Jacobi formulation. We have derived exact expression of the potential implicitly involved in  Mukhanov parametrization of inflationary equation-of-state for different values of the model parameter.  Using the latest available data for spectral index and tensor-to-scalar ratio we have imposed bounds on the model parameters.  Recent analysis from Planck \cite{akrami2020planck, aghanim2020planck}, WMAP and Bicep/Keck observations has set an upper-bound $r<0.036$ \cite{ade2021improved}  whereas it has been claimed that $r<0.032$  in Ref.\cite{tristram2022improved}. Throughout this work we have adhered to the following  constraints: $r<0.032$ and $0.9607\leq n_{_S}\leq 0.9691$.

We have also confronted the Mukhanov Parametrization with the futuristic ground based CMB mission CMB-S4 along with  another upcoming space mission LiteBIRD. We have found that  the non-detection of primordial gravity waves by CMB-S4 and  LiteBIRD may potentially rule out Mukhanov Parametrization along with many other inflationary models. However if these missions detect gravity waves then the model under consideration has a sufficient  parameter space where it can provide excellent fit to the observational data.

%
%
%

\section{Hamilton Jacobi Formalism }\label{hj}
The most commonly  used tool to solve the inflationary dynamics is the slow-roll approximations \cite{liddle1994}. But the need to go beyond slow-roll approximation is slowly but steadily licking in considering the ever increasing level of precision of the present  telescopes. To incorporate all the models irrespective of slow-roll approximations, Hamilton-Jacobi formalism \cite{muslimov1990, salopek1990}  has emerged as very powerful tool. Unlike slow-roll inflation, Hamilton-Jacobi formalism is more accurate as it does not ignore the contribution from the scalar kinetic term. The formulation is imitative by considering the inflaton field itself to be the evolution parameter. The key advantage of this formalism is that here we only need the Hubble parameter $H$, to be specified rather than the inflaton potential $V$. Since $H$ is a geometric quantity, unlike $V$, inflation is more naturally described in this language \cite{muslimov1990, salopek1990, lidsey1997}. Further, being first order in nature, these equations are easily tractable to explore the underlying physics.

The dynamics of a flat FRW unverse containing matter with density $\rho$ and  pressure $P$ is determined by the following equations 
\bea
H^2&\equiv&\left(\frac{\dot a}{a}\right)^2=\frac{1}{3{ M}_P^2}\rho\label{friedmann1}\\
\frac{\ddot a}{a}&=&-\frac{1}{6{ M}_P^2}\left(\rho+3P\right)\label{friedmann2}\\
{\dot\rho}&=&-3H\left(\rho+P\right)\label{cont}
\eea 
where $\rho$ and $P$ are energy and pressure densities respectively and ${\mbox{\rm M}}_P\equiv\frac{1}{\sqrt{8\pi G}}$ is the reduced Planck mass, $H=\dot{a}/a$ is the Hubble parameter and $a$ is the scale factor. Here an `over-dot' denotes derivative with respect to the cosmic time, $t$.  The energy and pressure densities of a homogeneous scalar
field with potential $V(\phi)$ are given by
\bea \label{inflaton:energypressure}
\rho&=&\frac{1}{2}{\dot\phi}^2+V(\phi), \ 
P=\frac{1}{2}{\dot\phi}^2-V(\phi).
\eea 
The Friedmann \eq{friedmann1} and \eq{friedmann2} for
the scalar field then turns out to be
\bea 
H^2&=&\frac{1}{3{ M}_P^2}\left[\frac{1}{2}{\dot \phi}^2+V(\phi)\right]\label{inflaton:friedmann1}\\
\frac{\ddot a}{a}&=&-\frac{1}{3{ M}_P^2}\left[\dot{\phi}^2-V(\phi)\right]\label{inflaton:friedmann2}
\eea
and the corresponding energy conservation turns out to be
\beq \label{inflaton:cont}
\ddot{\phi} + 3H{\dot\phi}+V'(\phi)=0 \eeq 
which is nothing but the Klein-Gordon equation for a scalar field. To derive the Hamilton-Jacobi equations in the context of inflation, we first take derivative of Eqn.\eqref{inflaton:friedmann1} with respect to cosmic time and 
substitute back into Eqn.\eqref{inflaton:cont} to obtain
\beq
2\dot{H}=-\mbox{ M}_P^{-2}\ \dot{\phi}^2.
\eeq
Now dividing both sides of the above equation by $\dot{\phi}$, we get
\beq\label{hj2}
\dot{\phi}=-2\mbox{ M}_P^2 H'(\phi)
\eeq
where a `prime' denotes derivative with respect to the scalar field $\phi$.
Finally substituting  Eqn.\eqref{hj2} back into Eqn.\eqref{inflaton:friedmann1} we get the following first order second degree non-linear differential equations \cite{salopek1990, muslimov1990,liddle1994, kinney1997,lidsey1997, barunquasi,videla2017, barun2018mutated, barunmhi, barunmhip}

\bea
\left[H^{\prime}(\phi)\right]^2 -\frac{3}{2M_P^2}
H(\phi)^2&=&-\frac{1}{2M_P^4}V(\phi)\label{qei:hamilton}\\
\dot{\phi}&=&-2M_P^2 H'(\phi).\label{qei:phidot}
\eea
The above two equations govern the inflationary
dynamics in Hamilton-Jacobi formalism. Now the acceleration  equation \eq{inflaton:friedmann2} may  be rewritten as 
\beq\label{adot}
\frac{\ddot{a}}{a}=-\left[2M_P^2H^{\prime}(\phi)^2-H(\phi)^2\right]\equiv H^2(\phi)\left[1-\rm\epsilon_{_H}\right] 
\eeq
where we have defined $ \epsilon_{_{H}}$ as
\beq\label{epsilon} \epsilon_{_{H}}=2 M_P^2\left(\frac{H'(\phi)}{H(\phi)}
\right)^2.
\eeq 
Inflation, i.e. accelerated expansion,  occurs when $\ddot{a}>0$  and this is uniquely determined by the condition $\epsilon_{_{H}}<1$ and inflation ends exactly at  $\epsilon_{_{H}}=1$. It is noteworthy to mention here  that the condition leading to a violation of the strong energy condition, $\rho+3P<0$, is uniquely determined by the magnitude of $\epsilon_{_H}$ alone. Now comparing \eq{adot} and \eq{friedmann2} we can easily derive the following relation \cite{garcia2014large, lidsey1997, kinney1997, kinney2002}
\bea 
\epsilon_{_{H}}&=&\frac{3}{2}(1+P/\rho)\equiv\frac{3}{2}(1+\omega)
\eea 
where $\omega\equiv P/\rho$, is the equation of state parameter. It is to be noted that $\epsilon_{_{H}}$ is not the usual  slow-roll parameter,  $\epsilon_{_{H}}$ measures the relative contribution of the inflaton's kinetic energy to its total energy \cite{liddle1994, lidsey1997, kinney1997,Copeland1998, kinney2002, garcia2014large} which is clear from the following relation
\bea
\epsilon_{_{H}}&\equiv&2M_P^2\left(\frac{H'(\phi)}{H(\phi)}\right)^2=\frac{\dot{\phi}^2}{2H^2}
\eea
It is customary to define another associated parameter, 
\begin{equation}\label{eta}
\eta_{_{H}}\equiv2M_P^2\frac{H''(\phi)}{H(\phi)}=- \frac{\ddot{\phi}}{H\dot{\phi}} 
\end{equation}
which measures the ratio of the field’s acceleration relative to the friction acting on it due to the expansion of the universe \cite{lidsey1997}. These two parameters are not the usual slow-roll parameters, but slow-roll approximation holds when these parameters are small in comparison to unity.

The amount of inflation is represented by number of e-foldings defined as 
\beq\label{efol}
N(t)\equiv \ln\frac{a(t_{\rm end})}{a(t)}=\int_{t}^{t_{\rm end}} H(t)dt
\eeq 
where $t_{\rm end}$ is the time when inflation comes to an end. We have defined $N$ in such a way that at the end of inflation $N=0$ and $N$ increases as we go back in time. The observable parameters are generally evaluated when there are $50-60$ e-foldings still left before the end of inflation. Though total number of e-foldings could be much larger. During this observable period inflationary EoS may be assumed very slowly varying or even almost constant. With the help of \eq{hj2} and \eq{epsilon}, the \eq{efol} can be rewritten as a function of the scalar field as follows
\beq\label{nphi}
N(\phi)=-\frac{1}{M_P^2}\int_{\phi}^{\phi_{\rm  end}}\frac{H(\phi)}{2H'(\phi)}\ d\phi=\frac{1}{M_P}\int_{\phi_{\rm  end}}^{\phi}\frac{1}{\sqrt{2\epsilon_{_{H}}}}\ d\phi=\int_{\phi_{\rm  end}}^{\phi}\frac{1}{\epsilon_{_{H}}}\frac{H'(\phi)}{H(\phi)}\ d\phi
\eeq
where $\phi_{\rm end}$ is the value of the scalar field at the end of inflation.

The above analysis of the Hamilton–Jacobi formalism assumes implicitly that the value of the scalar field is a monotonically varying function of cosmic time. In particular, it breaks down if the field undergoes oscillations. As a result, this formalism is not directly suitable for investigating the dynamics of a field undergoing oscillations in a minimum of the potential \cite{lidsey1997}. Also in practice the \eq{qei:hamilton} is very hard to solve due its non-linearity without invoking particular form of Hubble parameter. On the other hand once $H$ has been specified it is straight forward to anylyze Hamilton–Jacobi equations.
\section{Mukhanov Parametrization}\label{mukhanov}
Inflationary paradigm has been investigated recently through the parametrization of inflationary Equation-of-State (EoS henceforth) as a function of number of e-foldings, $N$. The main motivation was to develop a model independent framework for the investigation and confrontation of cosmic inflation with recent observations, and the  following  EoS parameter has been proposed in Ref.\cite{mukhanov2013quantum}
\beq\label{eos}
1+\omega=\frac{\beta}{({N}+1)^\alpha}
\eeq 
where $\alpha \ \mbox{and} \ \beta$ are two dimensionless positive constants and $N$ is the number of e-foldings. Above EoS captures a wide range of inflationary models with various observational predictions \cite{gariazzo2017primordial}. It has been argued   that with EoS parametrization, inflationary scenario can be studied model independently \cite{mukhanov2013quantum}. But this may not be true as we shall see that specifying a particular EoS indirectly translates into choosing a specific potential which has been  pointed out in Ref.\cite{martin2016observational}.

Since choosing an inflaton  potential directly maps into the underlying high energy physics, it is the most efficient way to learn about  inflation.   But nowadays highly sophisticated data are coming through various observational probes which are  leading the thoughts about cosmic inflation. With a specific EoS we might find inflationary observables more easily. So EoS formalism may be seen as bottom-up approach towards cosmic inflation.  Our aim here is to find the exact expression for the potential involved with the above EoS parametrization and set limit on the model parameters applying Hamilton-Jacobi formulation using latest constraints on inflationary observables.

\section{From EoS to Inflationary Potential}
Since inflation is a high energy phenomena, working with a  specific form of the potential directly establishes the link with high energy physics. Whereas EoS formalism helps us obtain observational predictions at ease without going into the details of underlying  physics. Now we shall see in the following analysis that within Hamilton-Jacobi formulation,  choosing an EoS is  equivalent to specify a particular Hubble parameter which in turn provides shape of the potential and vice versa.

Now, from the \eq{nphi} we can obtain 
\beq\label{hnphi}
dN=\frac{1}{\epsilon_{_{H}}}\frac{H'(\phi)}{H(\phi)}\ d\phi=\frac{1}{\epsilon_{_{H}}}\frac{dH}{H}.
\eeq
The integration of the above equation leads to the following form of the Hubble parameter 
\beq\label{hn}
{H(N)}={H_0}\exp\left(\int\epsilon_{_{H}}({N}){dN}\right) 
\eeq
where $H_0$ is the constant of integration. Now in the present context we have,  
\bea
\epsilon_{_H}=\frac{3\beta}{2({N}+1)^\alpha}\label{epsilon-n}
\eea
The end of inflation occurs naturally when $\epsilon_{_H}=1$ and from the definition of e-foldings  ${N}=0$ at the end of inflation. Considering these two we get a lower-bound of the parameter $\beta> 2/3$, since $\beta\leq 2/3$ would always imply $\epsilon_{_H}<1$ and inflation goes on forever. It is to be noted that graceful exit does not depend on the other free parameter, $\alpha$.

To obtain the evolution of the scalar field as a function of e-foldings, we now  rearrange the first equality of \eq{hnphi} and after squaring we get the following
\bea
\epsilon_{_{H}}&=& \frac{1}{2M_{P}^2}\left(\frac{d\phi}{dN}\right)^2\label{epphi},
\eea
which when combined with the \eq{epsilon-n} gives rise to 
\beq
d\phi=\sqrt{3\beta}{M_P}\frac{1}{(N+1)^{\alpha/2}}\ dN. 
\eeq
The above equation upon integration provides a relation between scalar field and number of e-foldings as follows,   
\beq
{\phi}=\left\{
\begin{array}{lll}C+\sqrt{3\beta}{M_P} \ln{(1+N)} & \mbox{if } \alpha =2 \\
	C+	\frac{\sqrt{3\beta}}{1-\alpha/2}{M_P}\left(1+{N}\right)^{1-\alpha/2}  & \mbox{if } \alpha\neq2 
\end{array}
\right.
\eeq
where $C$ is the constant of integration, which can be evaluated using the fact that at the end of inflation $\phi=\phi_{\rm end}$ and $N=0$. We have found  $C={\phi}_{\rm end}$ if $\alpha=2$ and $C={\phi}_{\rm end}-\frac{\sqrt{3\beta}}{1-\alpha/2}{M_P}$ if $\alpha\neq2$, after inserting these values of C into the previous equation we get
\beq\label{phi-n}
{\phi}=\left\{
\begin{array}{lll}
	{\phi}_{\rm end}+\sqrt{3\beta}{M_P}\ln{(1+N)} & \mbox{if } \alpha =2 \\
{\phi}_{\rm end}+	\frac{\sqrt{3\beta}}{1-\alpha/2}{M_P}\left(\left(1+{N}\right)^{1-\alpha/2} -1\right) & \mbox{if } \alpha\neq2 .
\end{array}
\right.
\eeq

Again we put  $\epsilon_{_H}$ from \eq{epsilon-n} into \eq{hn} and perform the integration to obtain 
\beq\label{hubble-n}
{H}=\left\{
\begin{array}{lll}
	{H}_0\left(1+{N}\right)^{\frac{3\beta}{2}}  & \mbox{if } \alpha =1 \\
	{H}_0\exp\left(\frac{3\beta}{2(1-\alpha)}\left(1+{N}\right)^{1-\alpha}\right)  & \mbox{if } \alpha\neq1 .
\end{array}
\right.
\eeq
Now from \eq{phi-n} and \eq{hubble-n} we determine the Hubble parameter in terms of the scalar field, 
\beq
{H}=\left\{
\begin{array}{lll}\label{hphi}
	{H}_0\left(2 \sqrt{3 \beta }\right)^{-3 \beta } (\phi/{M_P}) ^{3 \beta }  & \mbox{if } \alpha =1 \\
	{H}_0	\exp \left(-\frac{3\beta}{2}   \exp \left(-\frac{\phi }{\sqrt{3 \beta }{M_P}}\right)\right) & \mbox{if } \alpha=2\\
	{H}_0	\exp \left(-\frac{3 \beta \left(\frac{2-\alpha }{2 \sqrt{3 \beta }}\right)^{\frac{2-2 \alpha }{2-\alpha }}}{2 (1-\alpha )}\ (\phi/{M_P})^\frac{2-2 \alpha }{2-\alpha }\right)  & \mbox{if } \alpha\neq1,2
\end{array}
\right.
\eeq
Finally substituting   \eq{hphi} into \eq{qei:hamilton} we obtain the desired expression for the potential 
\beq\label{vphi}
V(\phi)=\left\{
\begin{array}{lllll}
	{M_P}^4	\ 2^{-6 \beta } 3^{1-3 \beta } \beta ^{-3 \beta } \left(\frac{\phi }{M_P}\right)^{6 \beta -2} \left(\left(\frac{\phi }{M_P}\right)^2-6 \beta ^2\right)& \mbox{if } \alpha =1 \\
	{M_P}^4\ \frac{3}{2} \left(2 e^{\frac{2 \phi\rm /M_P }{\left(\sqrt{3} \sqrt{\beta }\right)}}-\beta \right) \exp \left(-3 \beta  e^{-\frac{\phi \rm / M_P}{\left(\sqrt{3} \sqrt{\beta }\right) }}-\frac{2 \phi \rm /M_P }{\left(\sqrt{3} \sqrt{\beta }\right)}\right)& \mbox{if } \alpha =2 \\
	{M_P}^4\ \exp \left(\frac{2^{-\frac{2 (\alpha -1)}{\alpha -2}} 3^{\frac{1}{2-\alpha }} \beta  \left(-\frac{(\alpha -2) \phi }{\sqrt{\beta } M_P}\right)^{\frac{2 (\alpha -1)}{\alpha -2}}}{\alpha -1}\right)\left(3- \frac{2^{\frac{2-3\alpha}{\alpha -2}} 9^{\frac{1}{2-\alpha }} \beta ^2 \left(-\frac{(\alpha -2) \phi }{\sqrt{\beta } M_P}\right){}^{\frac{4 (\alpha -1)}{\alpha -2}}}{(\alpha -2)^2 \left(\frac{\phi }{\rm M_P}\right)^2}\right) & \mbox{if } \alpha\neq1,2.
\end{array}
\right.
\eeq
So the EoS \eq{eos} actually corresponds to the scalar field potential given by \eq{vphi}. Therefore we may conclude that specifying an EoS directly leads to a particular inflationary model as pointed out in \cite{martin2016observational}.

\section{From EoS to Inflationary Observables}
Considering ${N}$ as new time variable it is possible to express all the inflationary observables in terms of the EoS \cite{mukhanov2005, garcia2014large}, which up to the first order in slow-roll parameters are given by  
\bea
n_{_S} &\simeq&1 -3\left(1+\omega\right) + \dfrac{d}{d{N}}\ln\left(1+\omega\right)\\
\alpha_{_S} &\simeq& 3\dfrac{d}{d{N}}\left(1+\omega\right)-\dfrac{d^2}{d{N}^2}\ln\left(1+\omega\right)\\
r&\simeq&24\left(1+\omega\right)\\
n_{_T}&\simeq&-3\left(1+\omega\right)\\
\alpha_{_T} &\simeq&3\dfrac{d}{d{N}}\left(1+\omega\right)
\eea
where $n_{_T}$,  $\alpha_{_S}$, $\alpha_{_T}$ are tensor spectral index, running of scalar and tensor spectral indices respectively. 
To confront the prediction of any inflationary model the above observable quantities are evaluated at the time of horizon crossing i.e. when there are 50-60 e-foldings still left before the end of inflation. In this work we have  focused mainly on the scalar spectral index and the tensor-to-scalar ratio using their latest available bounds from recent observations \cite{ade2021improved,tristram2022improved}.  For the model under consideration we have 
\bea\label{r_ns}
n_{_S}&\sim& 1-\frac{3\beta}{(1+N)^{\alpha}}-\frac{\alpha}{(1+N)}\\
r&\sim& \frac{24\beta}{(1+N)^{\alpha}}\label{r_ns1}
\eea
From \eq{r_ns} and \eq{r_ns1} we can get the following relation between tensor-to-scalar ratio and scalar spectral index,
\beq
\frac{r}{8}\sim \left(1-n_{_S}\right)-\frac{\alpha}{(1+N)}.
\eeq
We shall later see that for $\alpha<1.50$ and  $\alpha>2.20$ the Mukhanov parametrization fails   to explain current observations. However when $1.5\leq\alpha\leq2.2$ the parametrization yields excellent match with observations. In the following section we have investigated  the model for different values of the model parameter $\alpha$.

\section{Inflationary Models}
\begin{figure}
	\begin{subfigure}[t]{0.5\textwidth}
		\includegraphics[width=9.5cm, height=5.9cm]{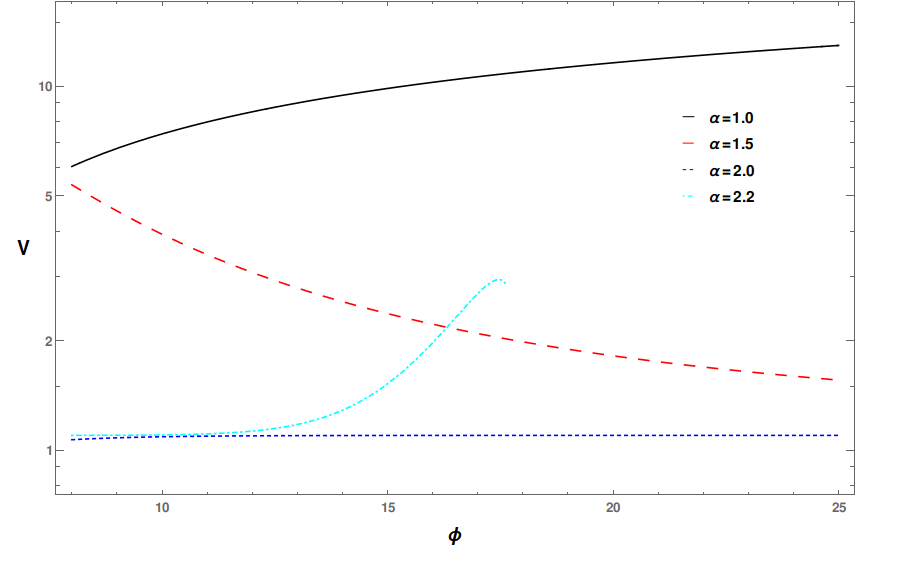}
		\caption{$\beta=1$}
	\end{subfigure}~
	\begin{subfigure}[t]{0.5\textwidth}
		\includegraphics[height=5.9cm]{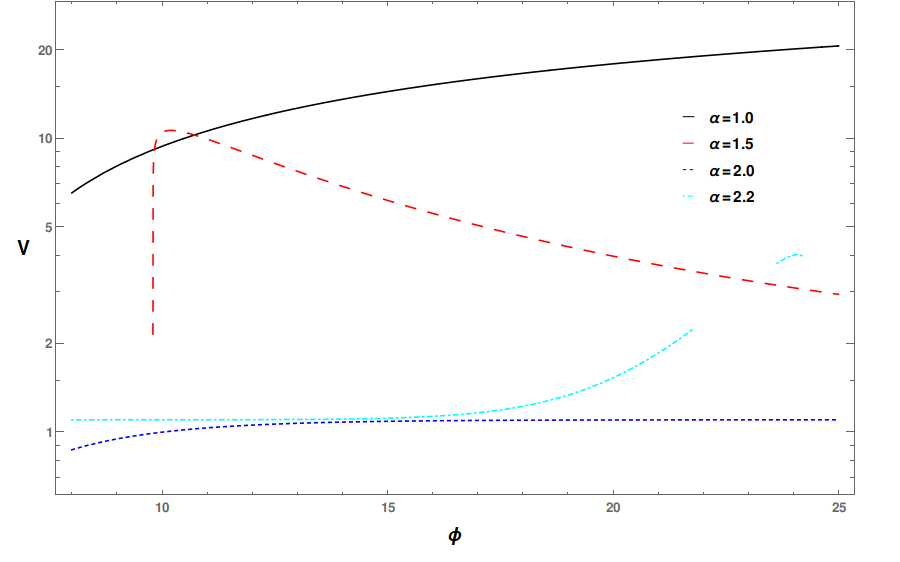}
		\caption{$\beta=2$}
	\end{subfigure}

	\begin{subfigure}[t]{0.5\textwidth}
	\includegraphics[width=9.5cm,height=5.9cm]{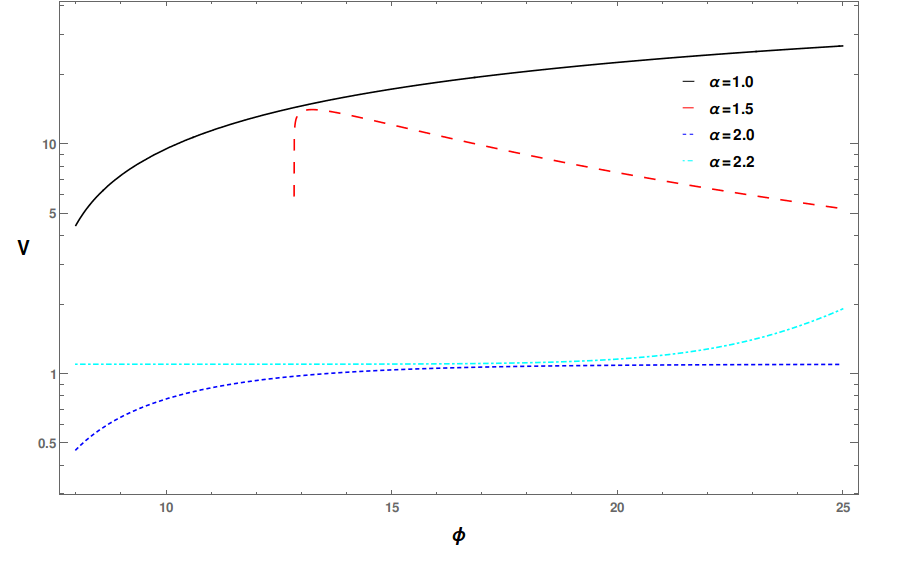}
	\caption{$\beta=3$}
\end{subfigure}~	
\begin{subfigure}[t]{0.5\textwidth}
	\includegraphics[height=5.9cm]{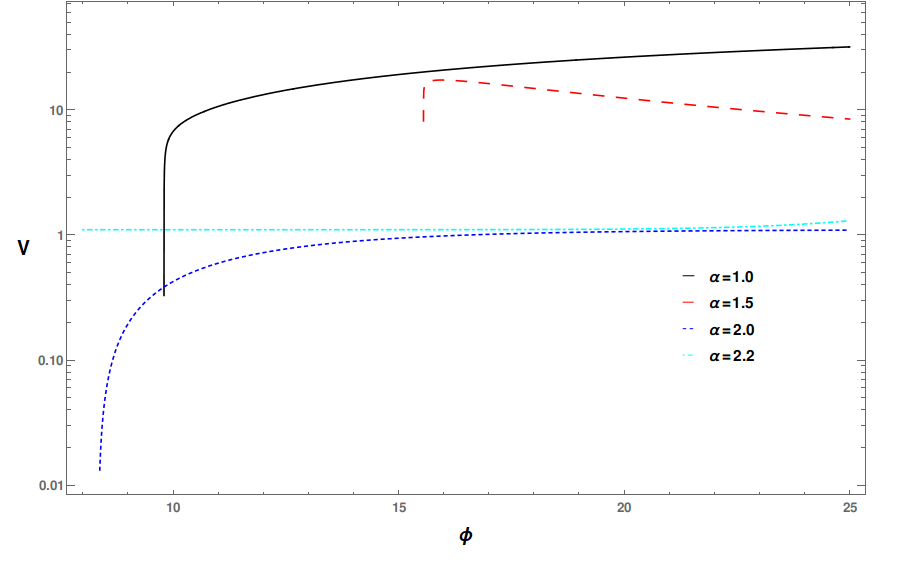}
	\caption{$\beta=4$}
\end{subfigure}

	\caption{\label{fig-vphi}Logarithmic variation of the potential with scalar field for four different values of the model parameter $\alpha$, $\beta$.}
\end{figure}

In this section we shall investigate few inflationary models by fixing the model parameter, $\alpha$.  In \fig{fig-vphi} we have plotted the inflationary potentials for different values of the model parameter $\alpha$ and for four values of  $\beta$ in logarithmic scale. From the figure it is very clear that the potential has adequate flat region which is  required in order to get sufficient amount of inflation to resolve big-bang puzzles. 

\subsection{Case $\alpha=1$}
For $\alpha=1$, the inflationary potential may  be put simply in the following  form
\bea
V(\phi)&=&2^{-6 \beta } 3^{1-3 \beta } \beta ^{-3 \beta } \left(\frac{\phi }{M_P}\right)^{6 \beta -2} \left(\left(\frac{\phi }{M_P}\right)^2-6 \beta ^2\right)\nonumber\\
&\equiv&V_0  \left[\left(\frac{\phi }{M_P}\right)^{6 \beta}-6 \beta ^2\left(\frac{\phi }{M_P}\right)^{6 \beta -2}\right]\label{vphi-1}
\eea
which represents power law potential and the corresponding inflationary scenario is the chaotic inflation \cite{linde1983, linde1983b}. The second term within the square brackets i.e., $-6 \beta ^2\left(\frac{\phi }{M_P}\right)^{6 \beta -2}$ differs from the original work in \cite{mukhanov2013quantum} as we are working in Hamilton-Jacobi formulation of inflation as it produces more accurate results. For $\beta=1/3$ \eq{vphi-1} represents quadratic potential whereas for $\beta=2/3$ it yields quartic potential. Later we shall show that both those values of $\beta$ are disfavoured as they produce higher amount of gravitational waves than that are observationally permitted.  

The spectral index and the tensor to scalar ratio in this case turn out to be 
\bea
n_{_S}&\sim& 1-\frac{3\beta+1}{1+N}\\
r&\sim& \frac{24\beta}{1+N}
\eea
\begin{figure}
	\centerline{\includegraphics[width=15.cm, height=10cm]{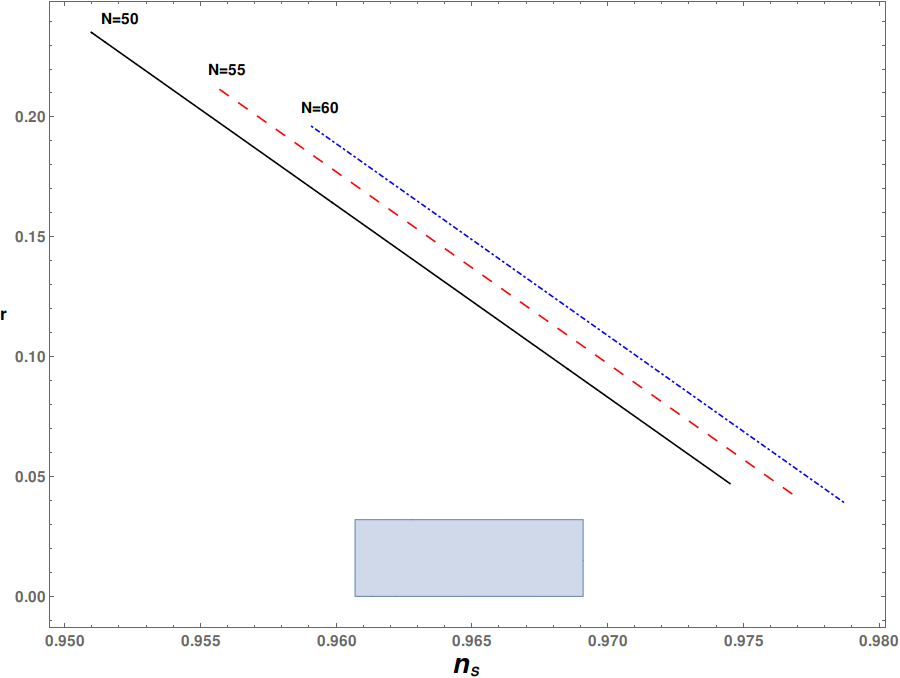}}
	\caption{\label{fig_rns1}Variation of the tensor-to-scalar ratio with the spectral index for 3 different values of number of e-foldings. The shaded region corresponds to the latest bounds on the spectral index and tensor-to-scalar ratio.  }
\end{figure}
From the observational constraint on the scalar spectral index, $0.9607\leq n_{_S}\leq 0.9691$, we find that $0.243\leq\beta\leq0.40$ for $N=55$. The lowest allowed value of $\beta=0.243$ comes up with tensor-to-scalar ratio of $0.10$ which is way too high for the recent findings of $r<0.032$ \cite{tristram2022improved}. 
So there is no feasible value of the model parameter $\beta$ which can simultaneously agree with present constraints on tensor-to-scalar ratio and scalar spectral index for $\alpha=1$. Which is also transparent from  \fig{fig_rns1}. The chaotic inflationary models with power law potential has already been ruled out due to the production of higher amount of primordial gravity waves than allowed observationally.   

\subsection{Case $\alpha=\frac{3}{2}$}
The inflaton potential in this case boils down to 
\beq
V(\phi)={M_P}^4 \ 3 e^{\frac{288 \beta ^2}{\left({\phi/M_P}\right)^2}} \left(1-55296 \frac{\beta ^4}{\left({\phi/M_P}\right)^6}\right)
\eeq

\begin{figure}
	\begin{subfigure}[t]{0.5\textwidth}
		\includegraphics[width=9.cm, height=5.9cm]{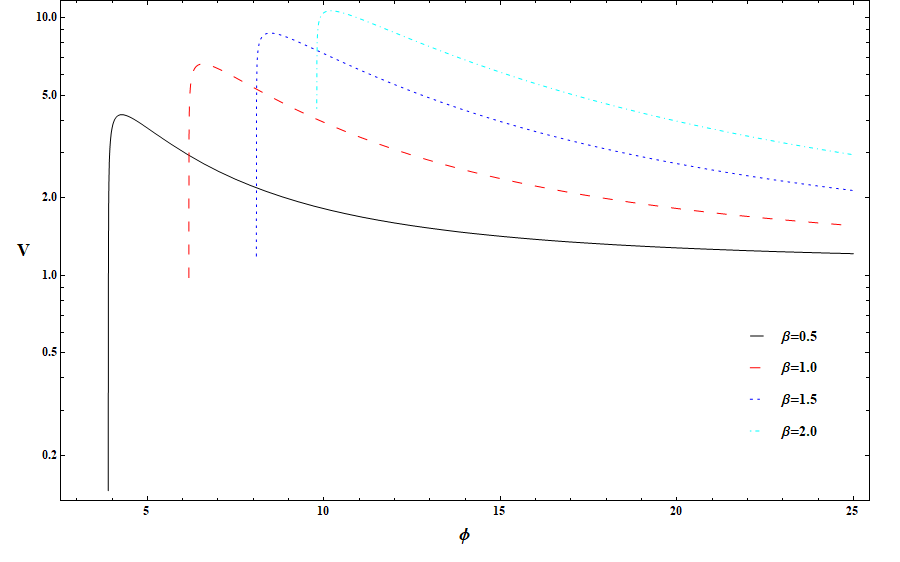}
	\end{subfigure}~
	\begin{subfigure}[t]{0.5\textwidth}
		\includegraphics[width=9.cm,height=5.9cm]{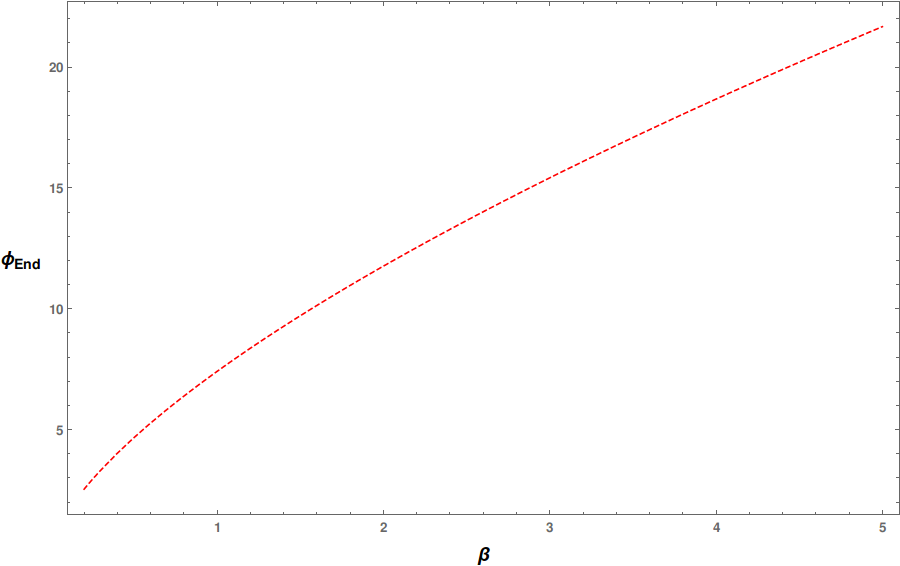}
	\end{subfigure}
	\caption{\label{v1.5}Left Panel: Logarithmic variation of the potential with scalar field for different values of $\beta$.\newline Right Panel: Variation of the end value of the inflaton with the model parameter $\beta$. }
\end{figure}
In \fig{v1.5} we have shown the logarithmic variation of the potential for three different values of the model parameter $\beta$. In this model inflation occurs at high values of the scalar field, which increases with the model parameter as can be seen from the right panel of the \fig{v1.5}. This high value of inflaton results in large  tensor-to-scalar ratio. In order to become observationally viable this type of inflationary model requires small values of the model parameter $\beta$. 

The spectral index and tensor-to-scalar ratio in this case are given by
\bea
n_{_S}&\sim& 1-\frac{3\beta}{(1+N)^{3/2}}-\frac{3}{2(1+N)}\\
r&\sim& \frac{24\beta}{(1+N)^{3/2}}
\eea
From the bound of tensor-to-scalar ratio, $r<0.032$, we can impose an upper bound on the model parameter which turns out to be $\beta<0.56$ for $N=55$. Also the observational constraint on scalar spectral index demands $0.575<\beta<1.75$ for $N=55$. 
\begin{figure}
	\begin{subfigure}[t]{0.5\textwidth}
		\includegraphics[width=9.cm, height=5.9cm]{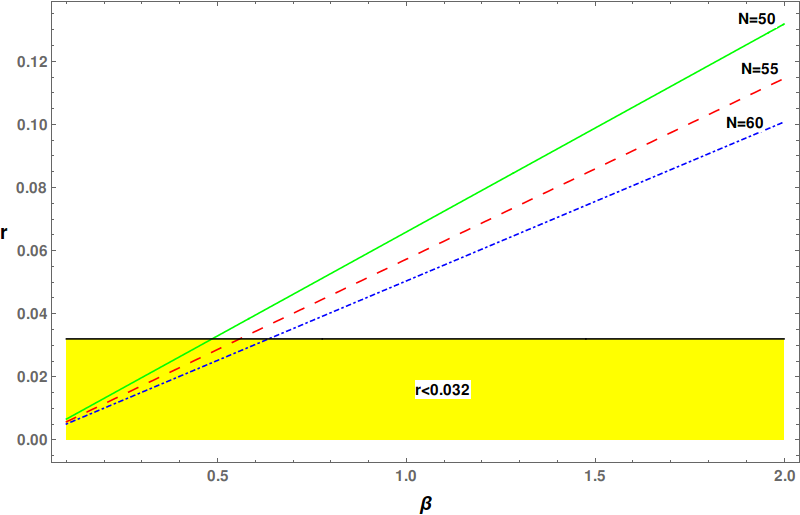}
	\end{subfigure}~
	\begin{subfigure}[t]{0.5\textwidth}
		\includegraphics[width=9.cm,height=5.9cm]{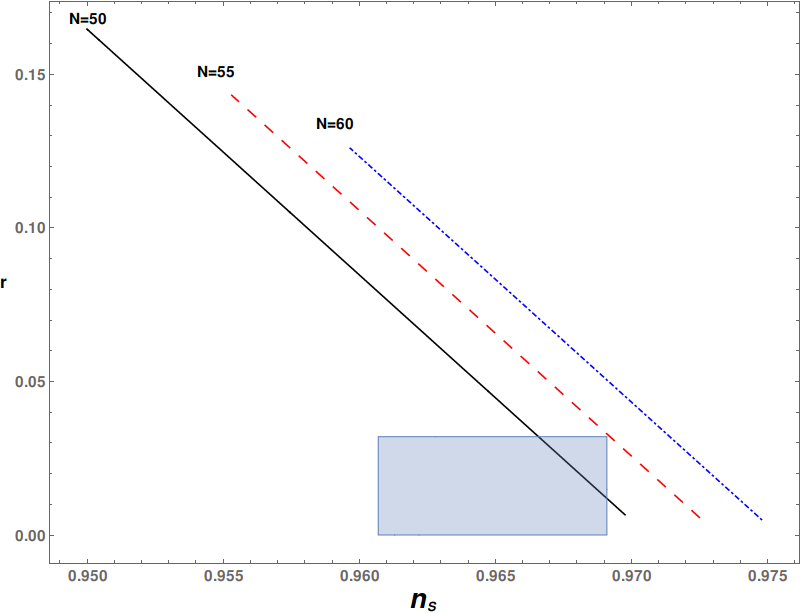}
	\end{subfigure}
	\caption{\label{rns_alpha15}Left Panel: Variation of the tensor-to-scalar ratio with the model parameter $\beta$. \newline Right Panel:Variation of the tensor-to-scalar ratio with spectral index for three different values of e-foldings $\beta$.}
\end{figure}
In the left panel of \fig{rns_alpha15} we have plotted tensor-to-scalar ratio for three different values of e-foldings, $N=50,55,60$, with the model parameter $\beta$. The shaded area refers  to observationally viable region.  In the right panel of \fig{rns_alpha15} we have shown the variation of tensor-to-scalar ratio with the scalar spectral index for three different values of e-foldings, $N=50,55,60$. The  shaded region corresponds to the $1-\sigma$ bound for spectral index  and $r<0.032$.

Here again we find that there is no feasible value of the model parameter, $\beta$, which is capable of explaining present bounds on tensor-to-scalar ratio and scalar spectral index simultaneously. But if we consider $N=50$, at the time of horizon crossing, we have a very small window where observational predictions for this model agree with the latest constraints. Since for  $N=50$, the observational constraint on tensor-to-scalar ratio implies $\beta<0.486$ while the $1-\sigma$  bound on spectral index demands $0.181<\beta<1.20$. So in this case the model parameter, $\beta$, has a tiny window,  $0.181<\beta<0.486$, where  observational bounds on   tensor-to-scalar ratio and scalar spectral index are met.  But then as $\beta<2/3$ inflation goes on forever and model suffers from graceful exit problem.  

\subsection{Case $\alpha=2$}
In this case, the inflationary potential may be written explicitly as 
\beq
V(\phi)=
{M_P}^4\ \frac{3}{2} \left(2 e^{\frac{2 \phi /M_P }{\left(\sqrt{3\beta }\right)}}-\beta \right) \exp \left(-3 \beta  e^{-\frac{\phi / M_P}{\left( \sqrt{3\beta }\right) }}-\frac{2 \phi  /M_P }{\left( \sqrt{3\beta }\right)}\right)
\eeq

\begin{figure}[!t]
	\begin{subfigure}[t]{0.5\textwidth}
		\includegraphics[width=9.cm, height=5.9cm]{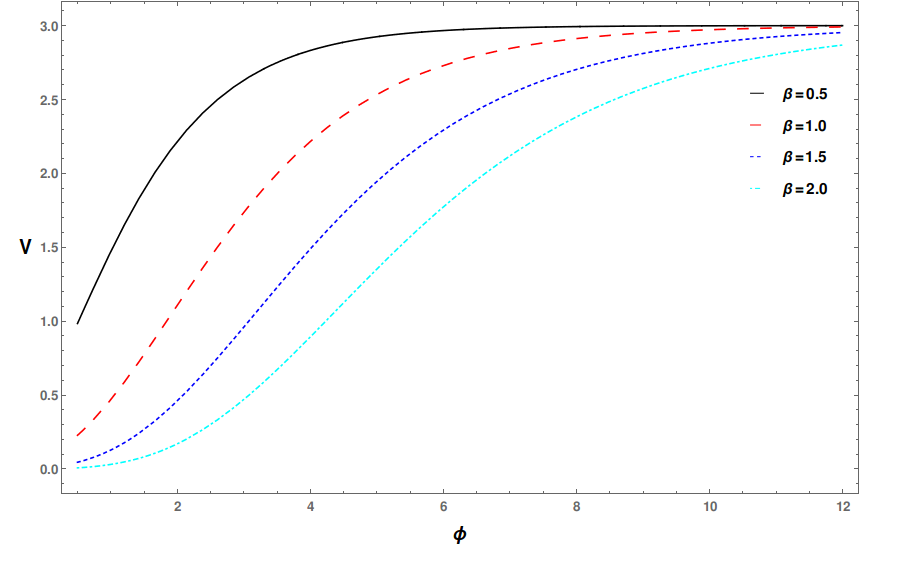}
	\end{subfigure}~
	\begin{subfigure}[t]{0.5\textwidth}
		\includegraphics[width=9.cm,height=5.9cm]{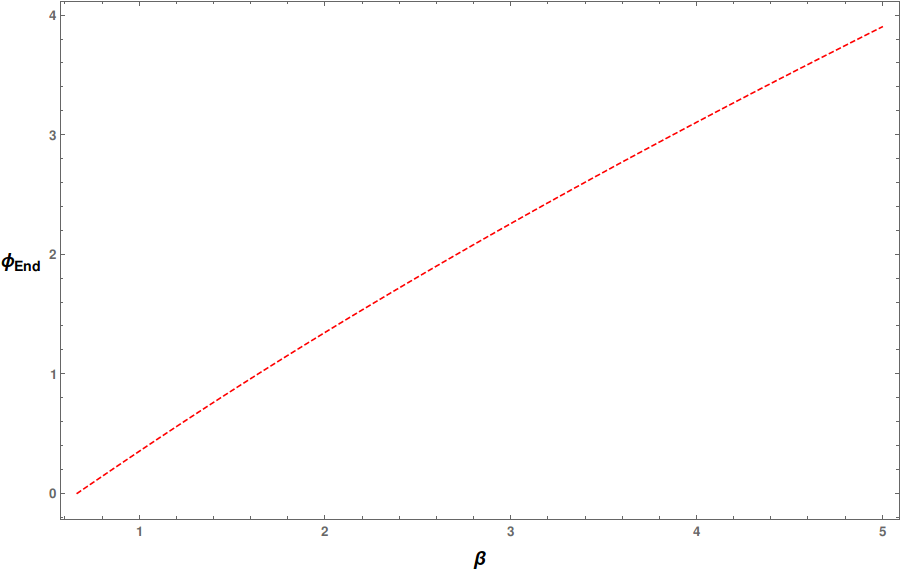}
	\end{subfigure}
	\caption{\label{v2}Left Panel: Variation of the potential with scalar field for different values of $\beta$.\newline Right Panel: Variation of the end value of the inflaton with the model parameter $\beta$. }
\end{figure}
\begin{figure}[!t]
	\begin{subfigure}[t]{0.5\textwidth}
		\includegraphics[width=9.cm, height=5.9cm]{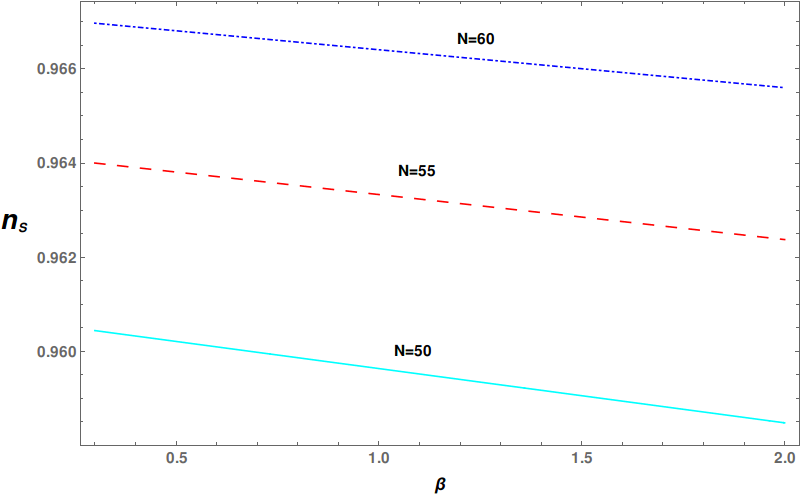}
	\end{subfigure}~
	\begin{subfigure}[t]{0.5\textwidth}
		\includegraphics[width=9.cm,height=5.9cm]{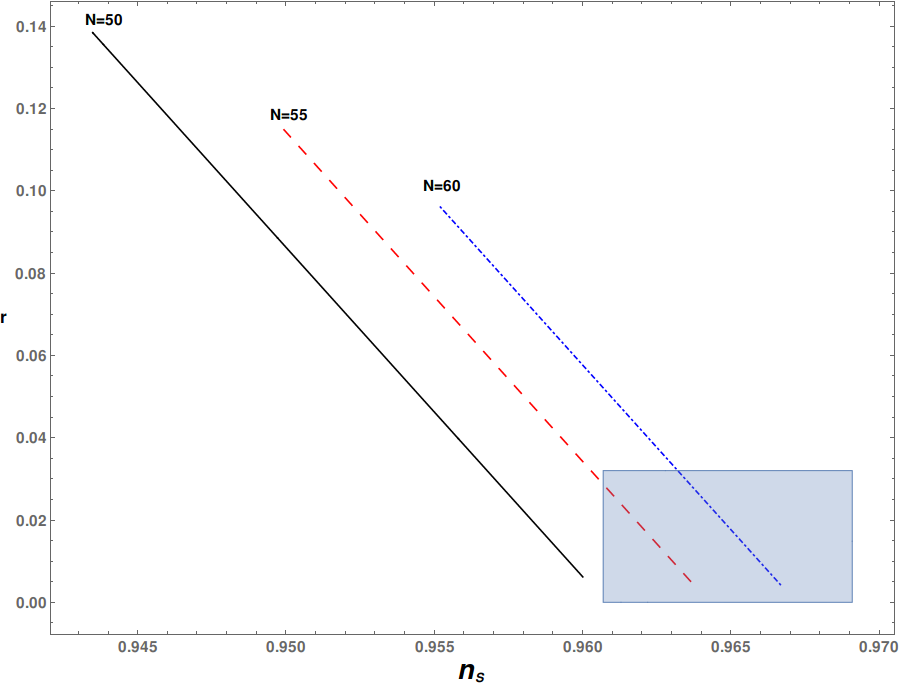}
	\end{subfigure}
	\caption{\label{fig_rns_alpha2}Left Panel: Variation of the spectral index  with scalar model parameter $\beta$.\newline Right Panel: Variation of tensor-to-scalar ratio with the scalar spectral index along with their observational bounds. }
\end{figure}

In the left panel of \fig{v2} we have shown the variation of the potential with inflaton for four different values of $\beta$ and in the right panel we have plotted the $\phi_{\rm end}$ with model parameter  $\beta$, which we have obtained by solving the equation $\epsilon_{_{H}}=1$. From the figure we notice that for $\beta<0.67$ we  have negative $\phi_{\rm end}$ which we are not interested in here and we shall only focus on the regime where  $\beta\geq0.67$. The scalar spectral index and the tensor-to-scalar ratio in this case turns out to be
\bea
n_{_S}&\sim& 1-\frac{3\beta+2+2N}{(1+N)^{2}}=1-\frac{2}{(1+N)}-\frac{3\beta}{(1+N)^{2}}\label{ns2}\\
r&\sim& \frac{24\beta}{(1+N)^{2}}\label{r2}
\eea
From the latest constraint on the spectral index we find that $\beta<3.75$ and that of tensor-to-scalar ratio implies $\beta<4.18$. So in this case the inflationary predictions are in tune with the latest available data for the model parameter within the range $0.67<\beta<3.75$. In the left panel of \fig{fig_rns_alpha2} we have shown the variation of the spectral index with the model parameter $\beta$ and we see that spectral index is almost independent of $\beta$ which is also clear from \eq{ns2}. In the right panel of \fig{fig_rns_alpha2} we have shown the variation of spectral index with the tensor-to-scalar ratio. The shaded region in \fig{fig_rns_alpha2} again refers to the observationally viable part. From the plot we clearly see that the model can simultaneously satisfy the constraints on tensor-to-scalar ratio and spectral index if we consider $N=55$.

\subsection{Case $\alpha>2$}
Finally we come to case $\alpha>2$ and the potential is precisely given  by
\beq
V(\phi)={M_P}^4\exp \left(\frac{2^{-\frac{2 (\alpha -1)}{\alpha -2}} 3^{\frac{1}{2-\alpha }} \beta  \left(-\frac{(\alpha -2) \phi }{\sqrt{\beta } M_P}\right)^{\frac{2 (\alpha -1)}{\alpha -2}}}{\alpha -1}\right)\left(3- \frac{2^{\frac{2-3\alpha}{\alpha -2}} 9^{\frac{1}{2-\alpha }} \beta ^2 \left(-\frac{(\alpha -2) \phi }{\sqrt{\beta } M_P}\right){}^{\frac{4 (\alpha -1)}{\alpha -2}}}{(\alpha -2)^2 \left(\frac{\phi }{\rm M_P}\right)^2}\right)
\eeq
\begin{figure}[!t]
	\begin{subfigure}[t]{0.5\textwidth}
		\includegraphics[width=9.cm, height=5.9cm]{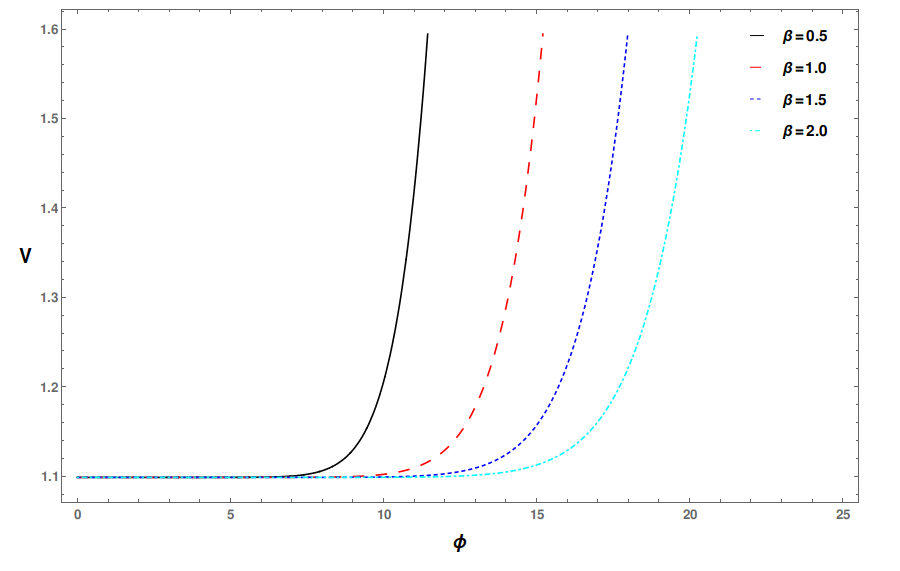}
	\end{subfigure}~
	\begin{subfigure}[t]{0.5\textwidth}
		\includegraphics[width=9.cm,height=5.9cm]{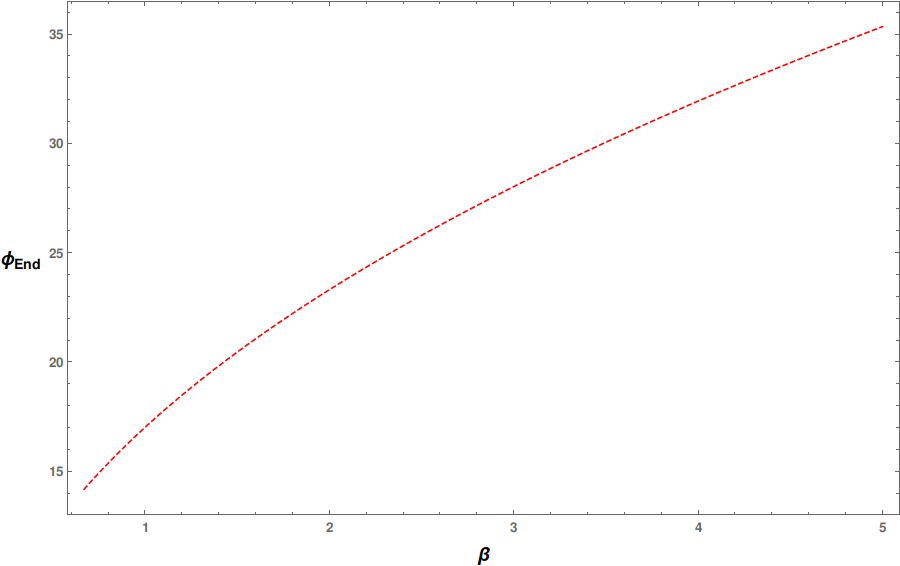}
	\end{subfigure}
	\caption{\label{v2.2}Left Panel: Variation of the potential with the scalar filed for 4 different values of the model parameter $\beta$ and for a fixed $\alpha=2.2$.\newline Right Panel: Variation of the end value of the inflaton with the model parameter $\beta$ for $\alpha=2.2$.}
\end{figure} 
In the left panel of \fig{v2.2} we have shown the logarithmic variation of the involved potential  with the scalar field. In the right panel we have plotted the variation of  $\phi_{\rm end}$ with the model parameter $\beta$.

\begin{figure}
	\centerline{\includegraphics[width=15.cm, height=10cm]{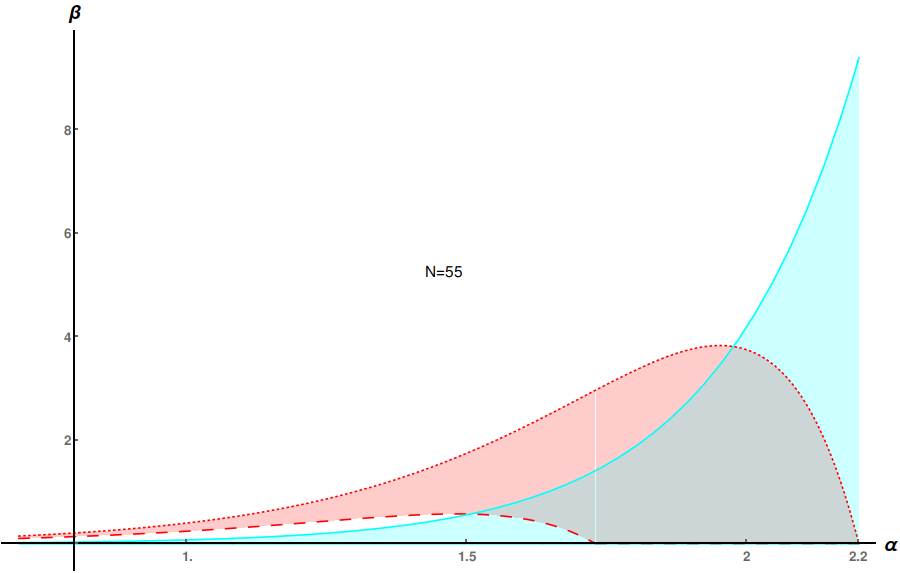}}
	\caption{\label{fig_beta_alphag55}Variation of the model parameters $\alpha$ and $\beta$. The red shaded region corresponds to the model parameters satisfying the latest constraint on the spectral index and the region with cyan represents models where constraint on tensor-to-scalar ratio is in tune.  }
\end{figure}

\begin{figure*}
	\begin{subfigure}[t]{0.5\textwidth}
	\includegraphics[height=6cm]{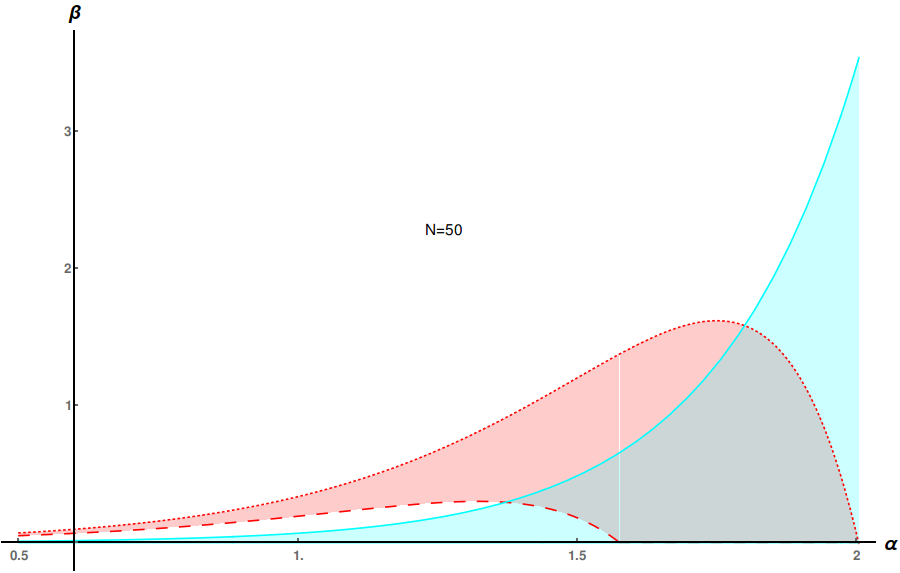}
		\caption{N=50}
	\end{subfigure}%
	~
	\begin{subfigure}[t]{0.5\textwidth}
		\includegraphics[height=6cm]{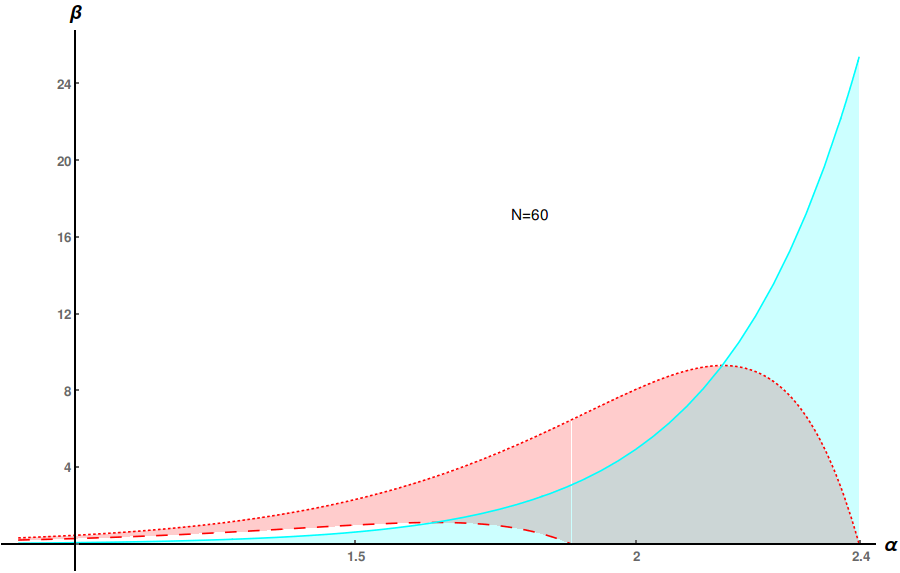}
		\caption{N=60}
	\end{subfigure}
	\caption{Variation of the model parameters $\alpha$ and $\beta$. The red shaded region corresponds to the model parameters satisfying the latest constraint on the spectral index and the region with cyan represents models where constraint on tensor-to-scalar ratio is in tune for $N=50$ and $N=60$ respectively.}\label{alpha-beta}
\end{figure*}

The expressions for the scalar spectral index and the tensor-to-scalar ratio are given by \eq{r_ns}. The observational constraint on tensor-to-scalar ratio, $r<0.032$, imposes following restriction on the model parameter $\beta$
\beq\label{r_beta_alphag2}
\beta < \frac{(1+N)^{\alpha}}{750}.
\eeq
Again from the above expression  of the spectral index \eq{r_ns} we obtain 
\beq
3\beta \simeq (1-n_{_S})\times(1+N)^{\alpha}-\alpha (1+N)^{\alpha-1},
\eeq
which implies that 
\beq\label{beta_alphag2}
\frac{1}{3}\left((1-n_{_S}|_{\rm upper})\times(1+N)^{\alpha}-\alpha (1+N)^{\alpha-1}\right)\lesssim \beta\lesssim\frac{1}{3} \left( (1-n_{_S}|_{\rm lower})\times(1+N)^{\alpha}-\alpha (1+N)^{\alpha-1}\right).
\eeq
where $n_{_S}|_{\rm upper/lower} $ corresponds to Planck-2018 $1-\sigma$ upper and lower bounds of the scalar spectral index respectively. Combining \eq{r_beta_alphag2} and \eq{beta_alphag2} we obtain the following upper-bound  of the model parameter
\beq
\beta< {\rm min}\left\{\frac{(1+N)^{\alpha}}{750},\frac{1}{3}\left((1-n_{_S}|_{\rm lower})\times(1+N)^{\alpha}-\alpha (1+N)^{\alpha-1}\right) \right\}
\eeq 
In \fig{fig_beta_alphag55} we have  shown the feasible region of the model parameter $\beta$ along with the other free parameter, $\alpha$. The region shaded in red corresponds to the values of $\beta$ which are in agreement with the latest bound of the scalar spectral index, whereas the region in cyan represents  values of $\beta$ satisfying current bound on tensor-to-scalar ratio for $N=55$. From \fig{fig_beta_alphag55} it is also obvious that the model is not capable of  explaining the inflationary observables whenever $\alpha\leq1.5$ or $\alpha\geq2.2$ for $N=55$. This bound obviously  depends on number of e-foldings. But whenever $1.5<\alpha\leq2.2$ we can easily find suitable values of $\beta$ which will simultaneously satisfy constraints  on scalar spectral index and tensor-scalar-ratio. In \fig{alpha-beta} we plotted the same as in \fig{fig_beta_alphag55} but now for  $N=50$ and $N=60$ respectively.

\begin{figure*}
	\begin{subfigure}[t]{0.5\textwidth}
		\includegraphics[height=6cm]{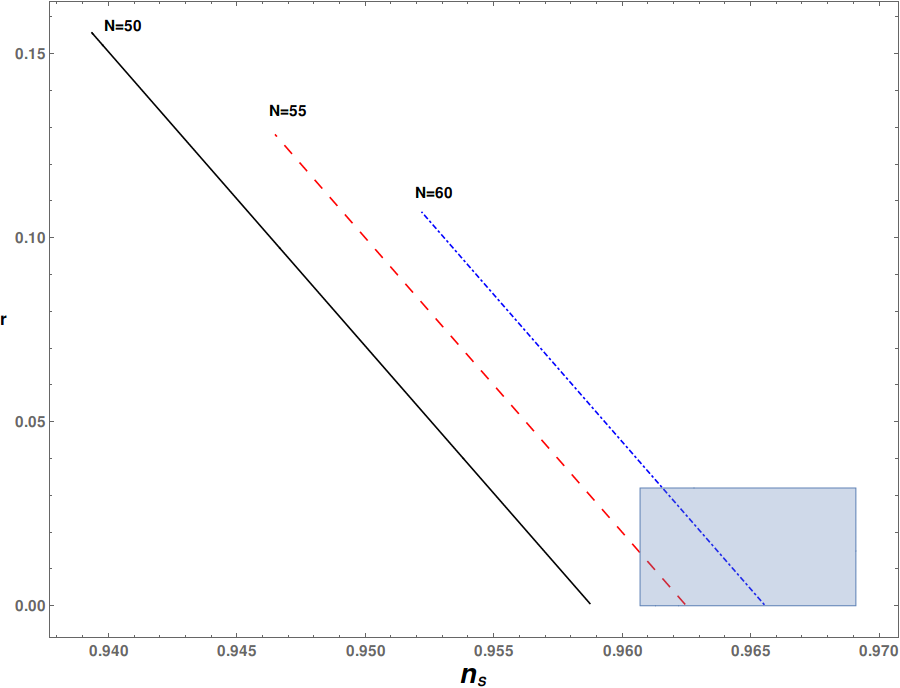}
		\caption{$\alpha=2.1$}
	\end{subfigure}%
	~
	\begin{subfigure}[t]{0.5\textwidth}
		\includegraphics[height=6cm]{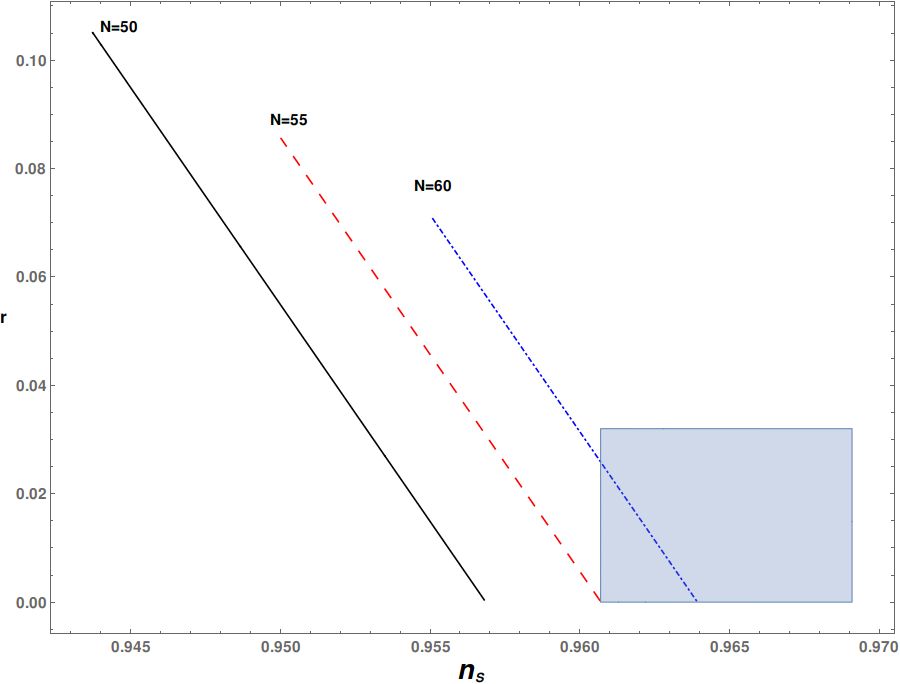}
		\caption{$\alpha=2.2$}
	\end{subfigure}
	\caption{\label{fig_rns2.2} Plot of tensor-to-scalar ratio with the scalar spectral index. The shaded region corresponds to the  latest constraint on the spectral index and tensor-to-scalar ratio.}
\end{figure*}

In \fig{fig_rns2.2} we have plotted the variation of tensor-to-scalar ratio with the scalar spectral index for $\alpha=2.1$ and 2.2. As before the shaded part refers to the $1-\sigma$ bound on the spectral index and $r<0.032$.

\section{Mukhanov Parametrization and Future CMB Missions}
Since the detection of gravity waves by LIGO and Virgo  scientific collaboration \cite{ligo2016a, ligo2016b}, we are eagerly waiting for the detection of large-scale primordial gravity waves through the CMB B-mode  polarization, which are supposed to be produced during inflation. The forthcoming ground based CMB-S4 \cite{abazajian2016cmb} mission along with space mission LiteBird \cite{matsumura2014mission} are expected to detect inflationary gravity waves provided $r>0.003$ or will set an upper-bound $r<0.001$ \cite{abazajian2022cmbS4,belkner2024cmbs4} in the absence of detection. In the context of upcoming CMB missions LiteBIRD and CMB-S4  we shall shed light on the parameter space for two free parameters, $\alpha, \ \beta$, in the following.
\subsection{LiteBIRD}
LiteBIRD, the Lite (Light) satellite for the study of B-mode polarization and Inflation from cosmic background Radiation Detection, is a space mission for primordial cosmology and fundamental physics.  The main goal of this mission is to measure tensor-to-scalar ratio, $r$, with uncertainty of $\delta r=0.001$ \cite{litebird2023probing, ghigna2024litebird}. A detection of primordial gravitational waves with LiteBIRD would shed light on the fundamental process occurred at an energy scale way beyond  the reach of present day particle accelerator.  A 3-$\sigma$ detection of gravity waves by the LiteBIRD would simply imply $r>0.003$ as it has sensitivity of $\delta r=10^{-3}$ within the multipole range $2\leq\ell\leq 200$. In the absence of detection of CMB B-mode polarization, LiteBIRD will set an upper bound $r<0.002$. 

To find the value of parameters $\alpha, \ \beta$, in the context of LiteBIRD we adhere to the following values of tensor-to-scalar ratio and scalar spectral index: $0.003<r<0.032$ and  $0.9607\leq n_{_S}\leq 0.9691$. Now using  the constraint on tensor-to-scalar ratio from the \eq{r_ns1} we can deduce that 
\beq
0.003<\frac{24\beta}{(1+N)^{\alpha}}<0.032\implies \frac{1}{8000}(1+N)^{\alpha}<\beta<\frac{1}{750}(1+N)^{\alpha}.
\eeq Also from the constraint on $n_{_S}$ and  \eq{r_ns} we get 
\beq
\frac{1}{3}\left(0.0309\times(1+N)^{\alpha}-\alpha (1+N)^{\alpha-1}\right)\lesssim \beta\lesssim\frac{1}{3} \left(0.0393\times(1+N)^{\alpha}-\alpha (1+N)^{\alpha-1}\right).
\eeq
\begin{figure}
	\centerline{\includegraphics[width=15.cm, height=10cm]{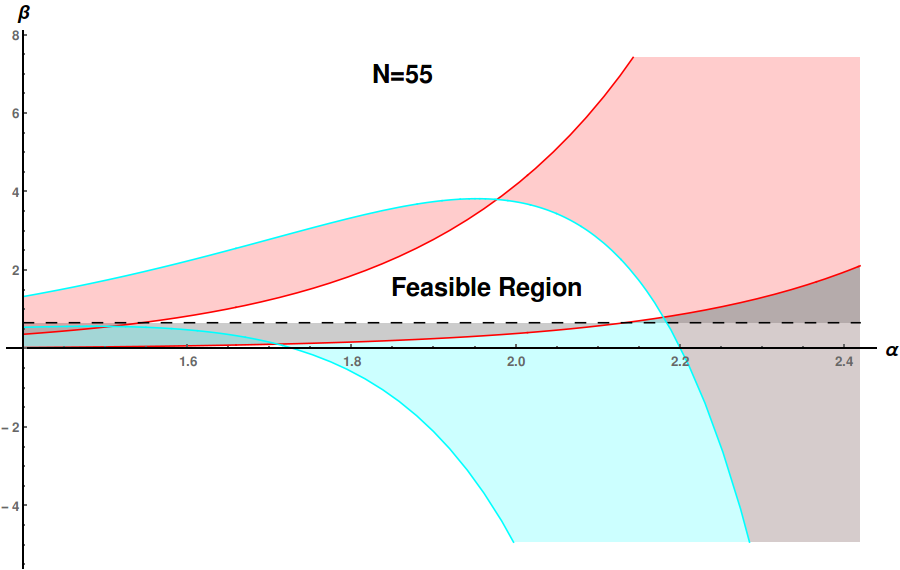}}
	\caption{\label{fig_beta_alphag55-lb-gw}Variation of the model parameters $\alpha$ and $\beta$. The  shaded region corresponds to the model parameters do not satisfy the latest constraint on the spectral index and tensor-to-scalar ratio.  The black dashed line corresponds to $\beta=2/3$, solutions below that line will suffer from graceful exit problem.}
\end{figure}
In \fig{fig_beta_alphag55-lb-gw} we have plotted the values of $\beta$ that are in accord with the scalar spectral index and tensor-to-scalar ratio with the other model parameter $\alpha$. The shaded region is the excluded area for the parameters  $\alpha$ and  $\beta$. The un-shaded region, marked as ``Feasible Region", where the observational predictions from Mukhanov Parameterization will be in accord with that of LiteBIRD. From the above figure one can easily conclude that if  $1.54387\leq\alpha\leq 2.17980$, then one can find suitable value of the model parameter $\beta$ so that EoS does mimic the observational predictions from LiteBIRD.       

If LiteBIRD fails to detect any primordial  gravity waves signal then the upper limit for the tensor-scalar-ratio will be pushed to  $r<0.002$,  which  translate into 
\beq
\beta< \frac{1}{12000}(1+N)^{\alpha},
\eeq
along with the following constraint from scalar spectral index: 
\beq
\frac{1}{3}\left(0.0309\times(1+N)^{\alpha}-\alpha (1+N)^{\alpha-1}\right)\lesssim \beta\lesssim\frac{1}{3} \left(0.0393\times(1+N)^{\alpha}-\alpha (1+N)^{\alpha-1}\right).
\eeq 
\begin{figure*}
	\begin{subfigure}[t]{0.5\textwidth}
		\includegraphics[height=6cm]{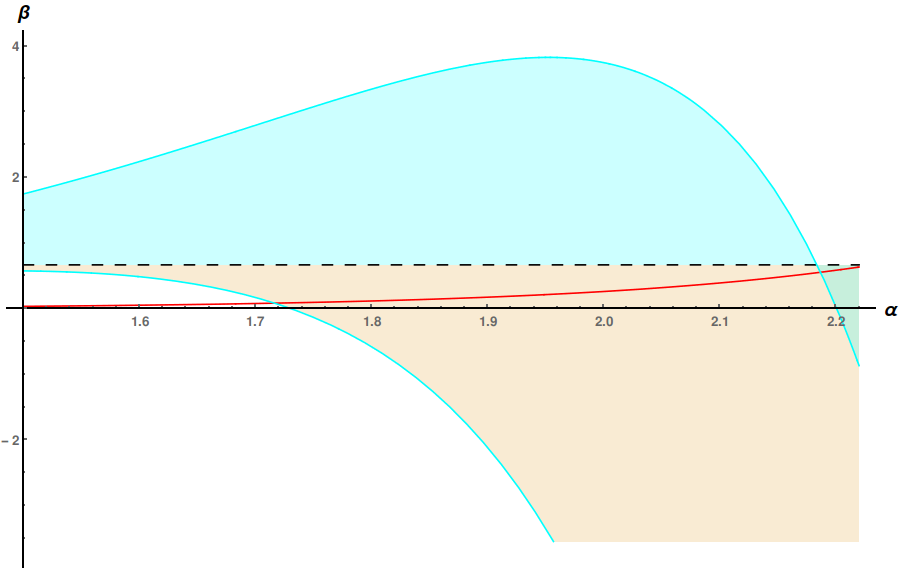}
		\caption{ LiteBIRD: N=55}
	\end{subfigure}%
	~
	\begin{subfigure}[t]{0.5\textwidth}
		\includegraphics[height=6cm]{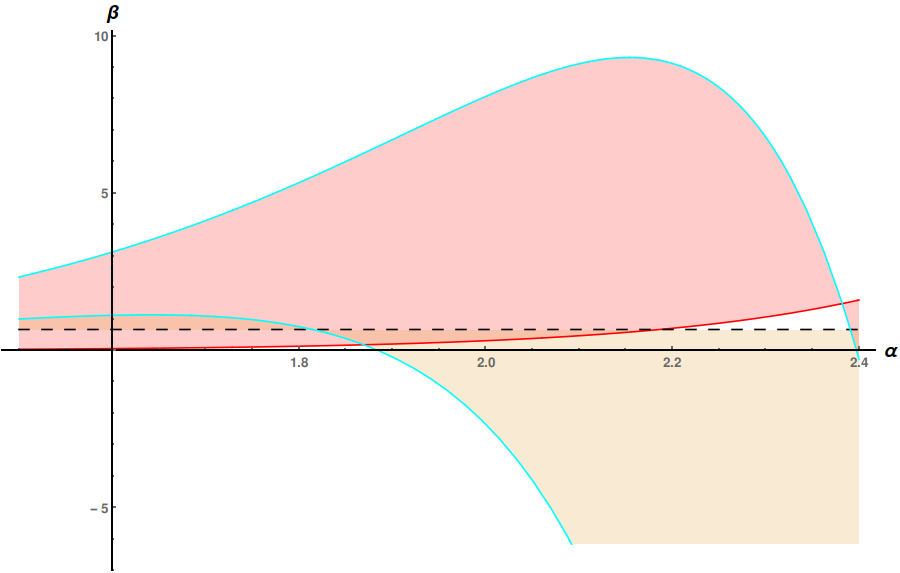}
		\caption{LiteBIRD: N=60}
	\end{subfigure}
	\caption{Variation of the model parameters $\alpha$ and $\beta$. The shaded region corresponds to the model parameters violating the observational constraints on the spectral index and tensor-to-scalar ratio for $N=55$ and $N=60$ respectively. The black dashed line corresponds to $\beta=2/3$, below that line solutions are not acceptable.}\label{lb-no-gw-alpha-beta}
\end{figure*}
In \fig{lb-no-gw-alpha-beta} we have shown the variation of $\beta$ with the model parameter $\alpha$ when LiteBIRD is unable to detect gravity waves. A very interesting outcome is that non detection of primordial gravity waves by LiteBIRD will completely rule out Mukhanov Parametrization. However, if one considers $N=60$, then one might get a tiny region  where we can have values of $\alpha$ and $\beta$ such that observational predictions are in tune with the model under consideration.

\subsection{CMB-S4}
\begin{figure}
	\centerline{\includegraphics[width=15.cm, height=10cm]{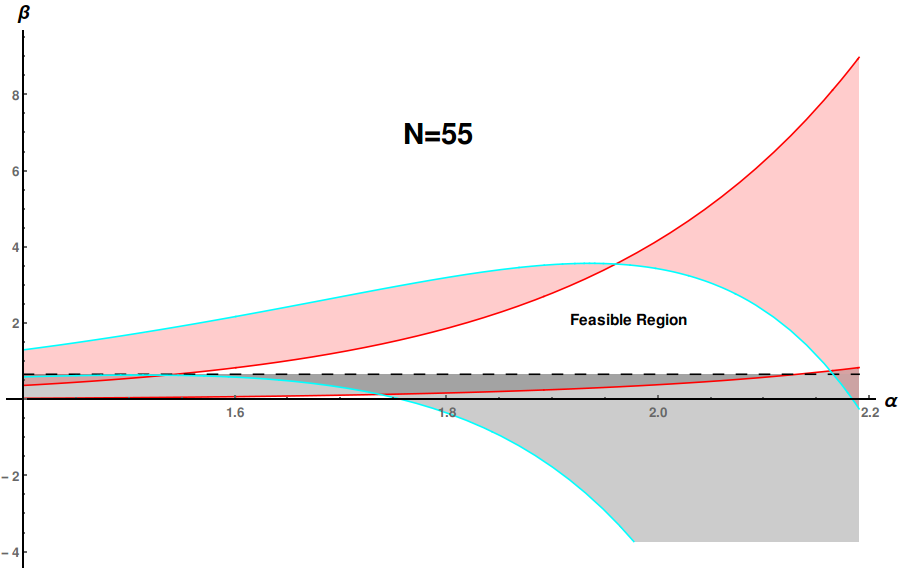}}
	\caption{\label{fig_beta_alphag55-s4-gw}Variation of the model parameters $\alpha$ and $\beta$. The  shaded region corresponds to the model parameters do not satisfy the latest constraint on the spectral index and tensor-to-scalar ratio.  The black dashed line corresponds to $\beta=2/3$. The region without shedding above the dashed line and below the solid cyan line represents feasible area.}
\end{figure}

A 5-$\sigma$ detection of primordial gravitational waves by  CMB-S4 would imply  $r>0.003$. And in this scenario  we find that Mukhanov Parametrization offers a wide range of parameter values where  constrains on tensor-to-scalar ratio and spectral index both are in accord with the observations. In \fig{fig_beta_alphag55-s4-gw} we have shown the feasible region ( non-shaded region above the black dashed line and below the solid cyan line) of the model parameters $\alpha,\ \beta$. 

The failure of CMB-S4 mission to observe primordial gravity waves would imply $r<0.001$, which when combined with \eq{r_ns1}  leads to  $\beta<\tfrac{1}{24000}(1+N)^\alpha$.  Also in order to avoid graceful exit problem we had earlier that $\beta>2/3$, which in combination  with the above constraint, $r<0.001$, leads to  an upper-bound for the other parameter, $\alpha>\frac{\ln 16000}{\ln(1+N)}$. So, $\alpha>2.46205, \  2.40484, \ 2.35481$ for $N=50,\ 55, \ 60$ respectively. But if $\alpha> 2.40484$ the constraint on spectral index is never satisfied for $\beta>2/3$ and $N=55$.  However if one assumes $N=60$, we have a tiny region where value of $n_{_S}$ is in accord with the CMB S-4 experiment. 
\begin{figure*}
	\begin{subfigure}[t]{0.5\textwidth}
		\includegraphics[height=6cm]{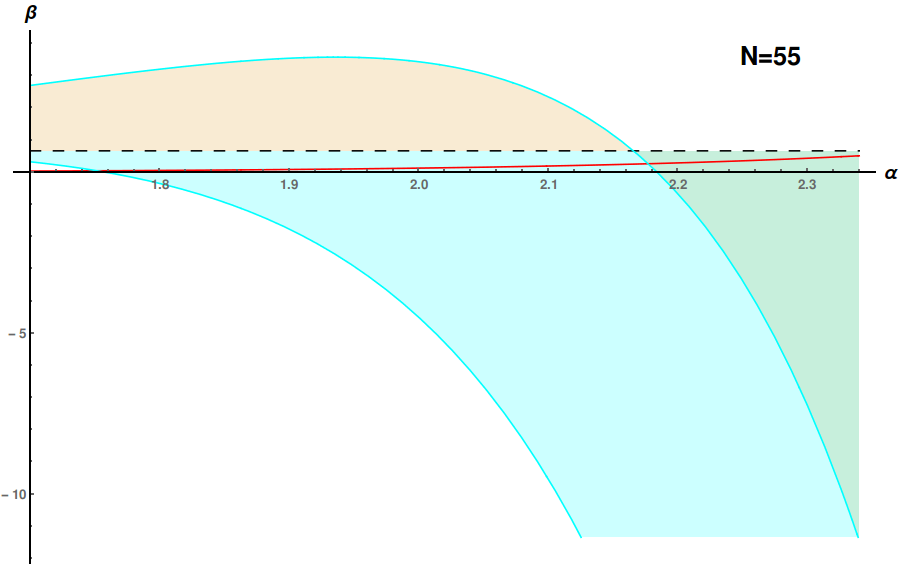}
		\caption{N=55}
	\end{subfigure}%
	~
	\begin{subfigure}[t]{0.5\textwidth}
		\includegraphics[height=6cm]{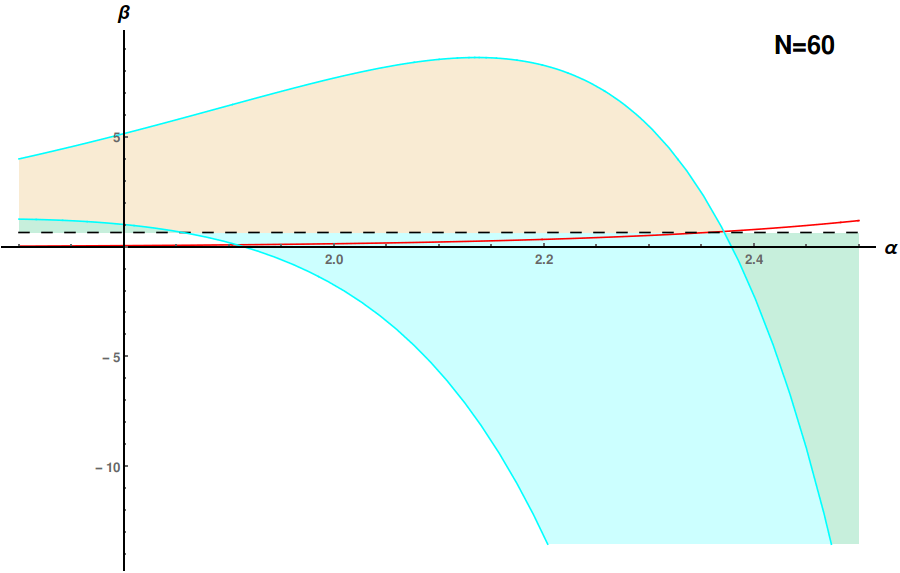}
		\caption{N=60}
	\end{subfigure}
	\caption{Variation of the model parameters $\alpha$ and $\beta$. The shaded region corresponds to the model parameters violating the observational constraints on the spectral index and tensor-to-scalar ratio  $N=55$ and $N=60$ respectively. The black dashed line corresponds to $\beta=2/3$.}\label{s4-no-gw-alpha-beta}
\end{figure*}
Hence non-detection of gravity waves by CMB S-4 experiment may rule out the Mukhanov Parametrization.

\section{Conclusion}
In this article we have reviewed the prospects and consequences of Mukhanov Parametrization of inflationary paradigm through the equation of state formalism employing Hamilton-Jacobi formulation. We have seen that the initial claim of model independent  study of inflation is not quite true, still EoS formalism can be very effectively utilized to describe early universe scenario. We find that the two parameters $\alpha$ and $\beta$ with numerical values close to unity can mimic the recent observations at ease.  In the process we have derived exact expressions for the potential which differs slightly from the original ones as we are working in the Hamilton-Jacobi formalism which is more accurate than the usual approach of slow-roll approximation.

We found that EoS as parametrized in \eq{eos} is in very good agreement with the recent observations for a decent range of the model parameters. Though we did not do rigorous numerical analysis but still able to put stringent constrains on the model parameters   $\alpha$ and $\beta$ using latest bounds on the inflationary observables. We found that whenever $1.5<\alpha\leq2.2$, we can  find suitable value of the other model parameter, $\beta$, so that the inflationary observables are in tune with the recent findings. In other words EoS formalism is capable of explaining the  latest observations  efficiently.  

The primordial gravity waves are yet be detected which will further restrict the spectrum of inflationary models. The ever improving accuracy of the spectral index and bound on the primordial gravity waves will certainly help to constrain the two model parameters $\alpha$ and $\beta$ further. Using the predictions from futuristic CMB missions in the likes of LiteBIRD and CMB-S4 experiments in order to obtain feasible region for the model parameters. We find that if LiteBIRD and  CMB-S4 detects primordial gravitational waves, then the parameter space for EoS will be shrunk. But if they are unable to detect the gravity waves then that might rule out Mukhanov parametrization for inflationary equation of state.

\section*{Acknowledgment } 
Author is grateful to the anonymous  reviewer of this work for his kind and useful suggestions. 
\begingroup
\renewcommand{\section}[2]{}%
\bibliography{references}
\endgroup

\end{document}